\documentclass[conference]{IEEEtran}
\usepackage[usenames,dvipsnames,svgnames,x11names]{xcolor}

\usepackage{cite}
\ifCLASSINFOpdf
  \usepackage[pdftex]{graphicx}
  \DeclareGraphicsExtensions{.pdf,.jpeg,.png}
\else
\fi
\usepackage{amsmath}
\usepackage{url}
\begin{document}
\title{Toward SATVAM: An IoT Network for Air Quality Monitoring}
\author{\IEEEauthorblockN{Rashmi Ballamajalu, Srijith Nair, Shayal Chhabra,\\
Sumit K Monga, Anand SVR,\\
Malati Hegde, Yogesh Simmhan}
\IEEEauthorblockA{Indian Institute of Science,\\Bangalore India\\
Email: brashmi@iisc.ac.in, malati@iisc.ac.in,\\
simmhan@iisc.ac.in}
\and
\IEEEauthorblockN{Anamika Sharma, Chandan M Choudhary,\\
Ronak Sutaria, Rajesh Zele}
\IEEEauthorblockA{Indian Institute of Technology, Bombay India\\
Email: rsutaria@iitb.ac.in, rajeshzele@ee.iitb.ac.in\vspace{0.5\baselineskip}}
\IEEEauthorblockN{Sachchida N. Tripathi}
\IEEEauthorblockA{Indian Institute of Technology, Kanpur India\\
Email: snt@iitk.ac.in}}

\maketitle

\begin{abstract}
Air pollution is ranked as the second most serious risk for public health in India after malnutrition. The lack of spatially and temporally distributed air quality information prevents a scientific study on its impact on human health and on the national economy. In this paper, we present our initial efforts toward \emph{SATVAM, Streaming Analytics over Temporal Variables for Air quality Monitoring}, that aims to address this gap. We introduce the multi-disciplinary, multi-institutional project and some of the key IoT technologies used. These cut across hardware integration of gas sensors with a wireless mote packaging, design of the wireless sensor network using 6LoWPAN and RPL, and integration with a cloud backend for data acquisition and analysis. The outcome of our initial deployment will inform an improved design that will enable affordable and manageable monitoring at the city scale. This should lead to data-driven policies for urban air quality management.
\end{abstract}


% \fi
\IEEEpeerreviewmaketitle

\section{Introduction}
\IEEEPARstart{U}rban air pollution is one of the biggest health challenges of India as the country rapidly urbanizes in an unstructured manner. While vehicular pollution is one of the visible contributors, pollution from industries and from burning of crop residue are also leading causes. The key pollutants as identified by the Central Pollution Control Board (CPCB) of India are: PM\textsubscript{2.5}, PM\textsubscript{10}, NO$_2$, SO$_2$, O$_3$, CO besides other VOCs and heavy metals \cite{cpcb_2009}. The systematic study of air pollution in India has been constrained due to the limited number of reference air quality monitors available in the field \cite{guttikunda_2017}.
Low-cost real-time GPRS (General Packet Radio Service) based air quality monitoring networks have been launched in India by non-profit groups~\cite{breathe_2016} with evaluation and calibration of PM\textsubscript{2.5} sensors in the field~\cite{amt-2018-111}. However, these need to scale much more with improved affordability and robustness. 
As a result, an effective development of data-driven pollution management policy is currently infeasible.

\emph{SATVAM -- Streaming Analytics over Temporal Variables from Air quality Monitoring} -- is a recently initiated project to develop tools, technologies, architectures and analytics to deploy and study air quality monitoring at city-wide scales and at finer spatial and temporal granularity than currently done. Funded by Department of Science and Technology (DST) and Intel, and administered by the Indo-US Science and Technology Forum (IUSSTF)~\footnote{IUSSTF Real Time River Water \& Air Quality Monitoring, http://www.iusstf.org/program/real-time-river-water--air-quality-monitoring}, this project aims to develop an IoT network for air quality monitoring that leverages four research vectors: low-cost sensing, energy harvesting, low-power communication networks, and edge analytics. The project is led by IIT Kanpur and includes partners from the Indian Institute of Science (IISc) and IIT Bombay from India, and Duke University from the US.

In this article, we examine the initial design, integration and small-scale deployment of Gas Sensor Nodes connected through Low Power Wide Area Network (LP-WAN), and using edge devices for basic analytics and data acquisition for subsequent visualization in the cloud. This setup will be deployed in New Delhi to collect reference data from our commodity sensors, co-located with higher quality sensors at the Indian Meteorological Department. This serves to integrate and validate the constituent technologies, and offer initial data for calibration of the commodity sensors. We present our gas sensor hardware integration in Sec.~\ref{sec:hardware}, wireless sensor network design in Sec.~\ref{sec:design}, and offer our conclusions and future plans for eventual larger scale deployment in Sec.~\ref{sec:conclude}.

%%%%%%%%%%%%%%%%%%%%%%%%%%%%%%%%%%%%%%%%%%%%%%
\section{Gas Sensor Node Design}
\label{sec:hardware}
%The edge sensor nodes for Real-time Air Quality Monitoring developed here
IoT has been a key enabling architecture for smart cities ~\cite{simmhan:spe:2018}. This project aims to create a scalable IoT infrastructure that can monitor and analyze air pollution levels in an outdoor environment. Low cost sensors measure certain harmful gases such as NO$_2$, SO$_2$, O$_3$ and particulate matter categorized as PM\textsubscript{2.5}, PM\textsubscript{10} present in the air. 
Our architecture uses a Wireless Sensor Network (WSN) based on the \emph{Zolertia RE-Mote (Rev. B)}~\cite{re-mote} for communication, and gas sensors are integrated with the mote for sensing air quality.
The four-electrode B4 series low-power electrochemical toxic gas sensor modules from \emph{Alphasense} are used. The gas sensors are designed for environmental application, and when exposed to ambient air, measure the target gas concentration. 

\begin{figure}[t]
\centering
\includegraphics[width=0.90\columnwidth]{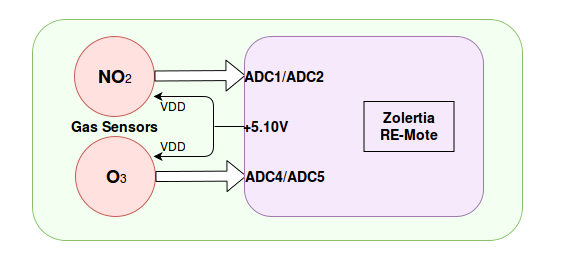}
\vspace{-0.2in}
\caption{Gas sensor integration with RE-Mote}
\label{fig:sens_int}
\vspace{-0.1in}
\end{figure}

The block diagram of the sensor integration with the RE-Mote is shown in Fig.~\ref{fig:sens_int}.  The RE-Mote is used as the Analog to Digital Converter (ADC) to interface with the analog sensor data. The RE-Mote consists of CC2538 Wireless Micro-controller SoC~\cite{cc2538} which has an internal 12-Bit ADC with 8 analog input channels. The gas sensor module outputs potential of Working Electrode and Auxiliary Electrode potential in voltage (millivolts). The sensors work at the VDD range of 3.5~to 6.4~VDC. The gas sensors are powered up using the +5.10V DC output pin of the RE-Mote. The analog outputs of the gas sensor module are labeled as OP1 and OP2 for Working electrode and Auxiliary Electrode respectively. These analog sensor output pins are interfaced to the analog input channels of the RE-Mote. The analog sensor output are converted to digital data using the internal ADC of RE-Mote. The digital output values from the ADC are obtained in mV as:
\[ Analog\_op = Digital\_op \times \frac{3300}{4095} \]
where $3300~mV$ corresponds to the internal voltage reference of the ADC corresponding to its $4096$ discrete levels. The analog data obtained from the RE-mote is validated using digital multimeter measured directly at the sensor OP pins.

The current implementation includes integration of two toxic gas sensors, NO$_2$ and O$_3$ (ozone), connected to four ADC channels of the RE-Mote. Channels 1 and 2 are used for the NO$_2$ sensor output while ADC channels 4 and 5 are used for the O$_3$ sensor. Besides these two gas sensors, we plan to integrated SO$_2$ gas sensors over ADC, and PM2.5, PM10 over UART with the RE-mote in the near future for holistic monitoring.

%%%%%%%%%%%%%%%%%%%%%%%%%%%%%%%%%%%%%%%%%%%%%%

\section{IoT and Wireless Network Design}
\label{sec:design}
 Multiple gas sensor devices are connected through a wireless mesh network infrastructure, that is supported by a wired backhaul network. The initial deployment consists of the Zolertia RE-motes operating in the Sub-1~GHz frequency, and forming a \emph{6LoWPAN network (IPv6 over Low-Power Wireless Personal Area Networks)}~\cite{rfc4944} with a \emph{border router} using \emph{RPL (Routing Protocol for Low-Power and Lossy Networks)}~\cite{rfc6550} to collect the data over the wireless medium. 

\begin{figure}[t]
\centering
\includegraphics[width=1.0\columnwidth]{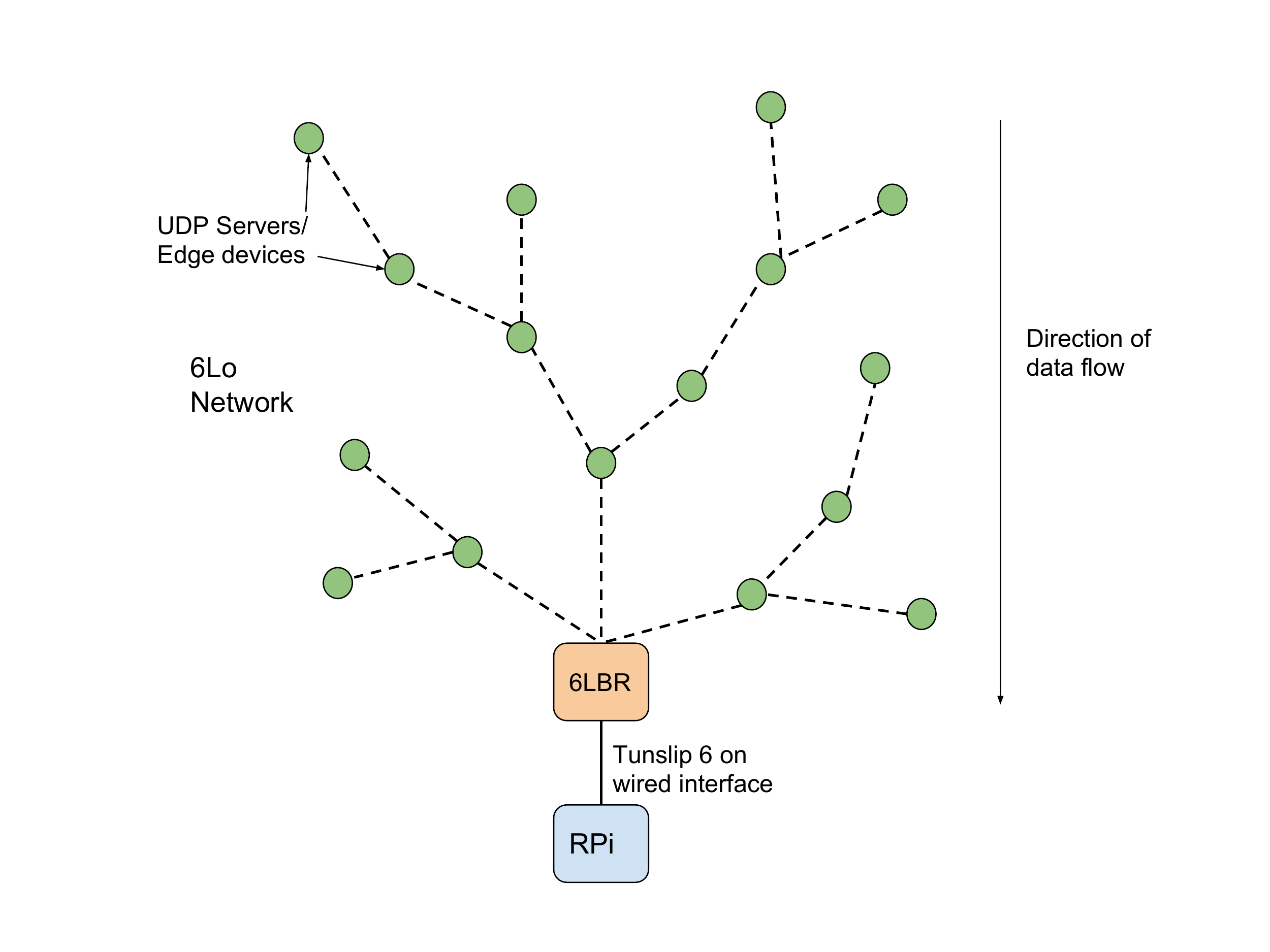}
\vspace{-0.4in}
\caption{Tree structure of the 6LoWPAN network with RPi gateway}\label{fig:lpwan-tree}
\end{figure}

RPL is used to form a Directed Acyclic Graph (DAG) for routing packets in the network, from the edge devices containing the sensors to the border router. The overall structure of the network is that of a tree with the border router forming the root of the DAG, the edge devices forming the leaf nodes and/or intermediate relay nodes.  The border router is in turn connected to the backhaul network through a Raspberry Pi that sends data over a private broadband network to the cloud. An example setup is shown in Fig~\ref{fig:lpwan-tree}. This wired backhaul will be replaced with a GSM or Narrowband IoT (NB-IoT) network for future deployments. Next, we offer details on this overall network architecture.

\begin{figure}[t]
\centering
\includegraphics[width=1.0\columnwidth]{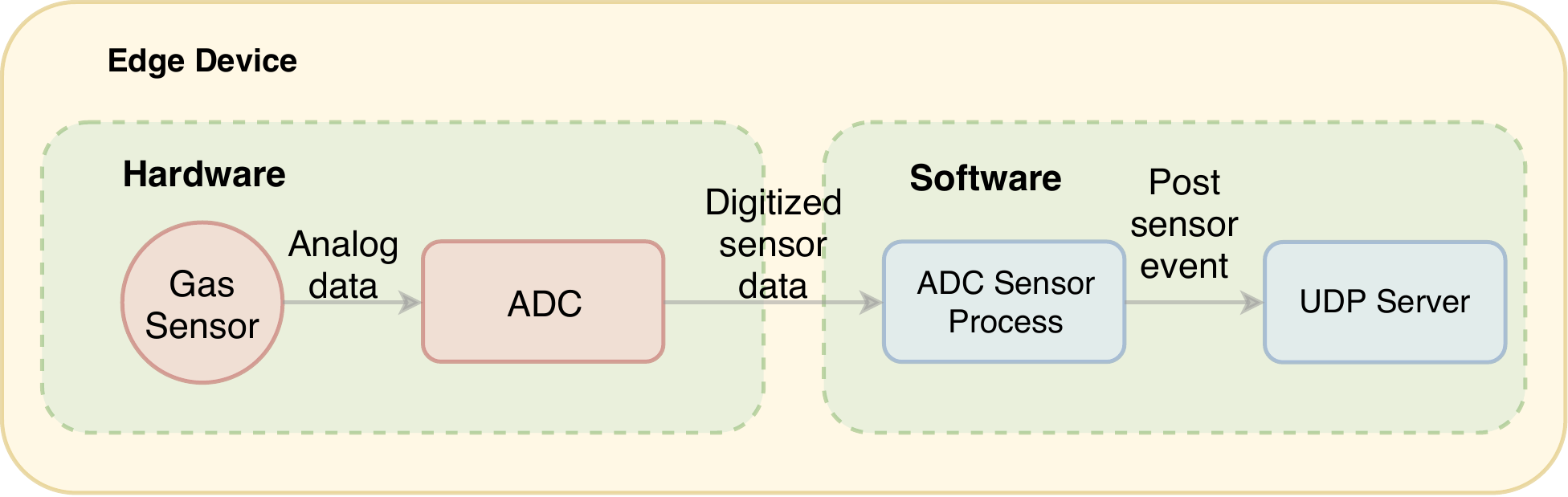}
\vspace{-0.2in}
\caption{Process Flow at RE-Mote edge device}\label{fig:mote-flow}
\vspace{-0.1in}
\end{figure}

\textbf{Sensor Data Acquisition:} The sensor data at the ADC ports can be monitored continuously and instantaneously, based on the application requirements. The sensor data is currently sampled with a frequency of 0.25 samples/second, wherein each sample consists of one sensor data element. The current implementation includes NO$_2$ and O$_3$ sensors connected to 4 ADC channels on RE-mote, as discussed earlier, with each sensor providing 2 output parameters of size 4 bytes each. We are also including a sampling counter of size 4 bytes, for the purpose of unique identification of packets in the cloud. The overall data from the two sensors, thus, adds up to 20 bytes.

Our network stack is implemented on the \emph{Contiki} operating system~\cite{contiki}, which is a lightweight operating system for low power IoT devices and runs on the RE-mote.  
The ADC sensor process is implemented as a \emph{process thread} in Contiki created with a \emph{process ID, the type of event} and \emph{the data payload} generated by it. When this process samples data from the ADC, an event is created and posted to a listener \emph{UDP server process}, as shown in the Fig~\ref{fig:mote-flow}. The OS kernel, using its internal event queue dispatch mechanism, delivers the ADC event data to the UDP server process thread.

\begin{figure}[t]
\centering
\includegraphics[width=1.0\columnwidth]{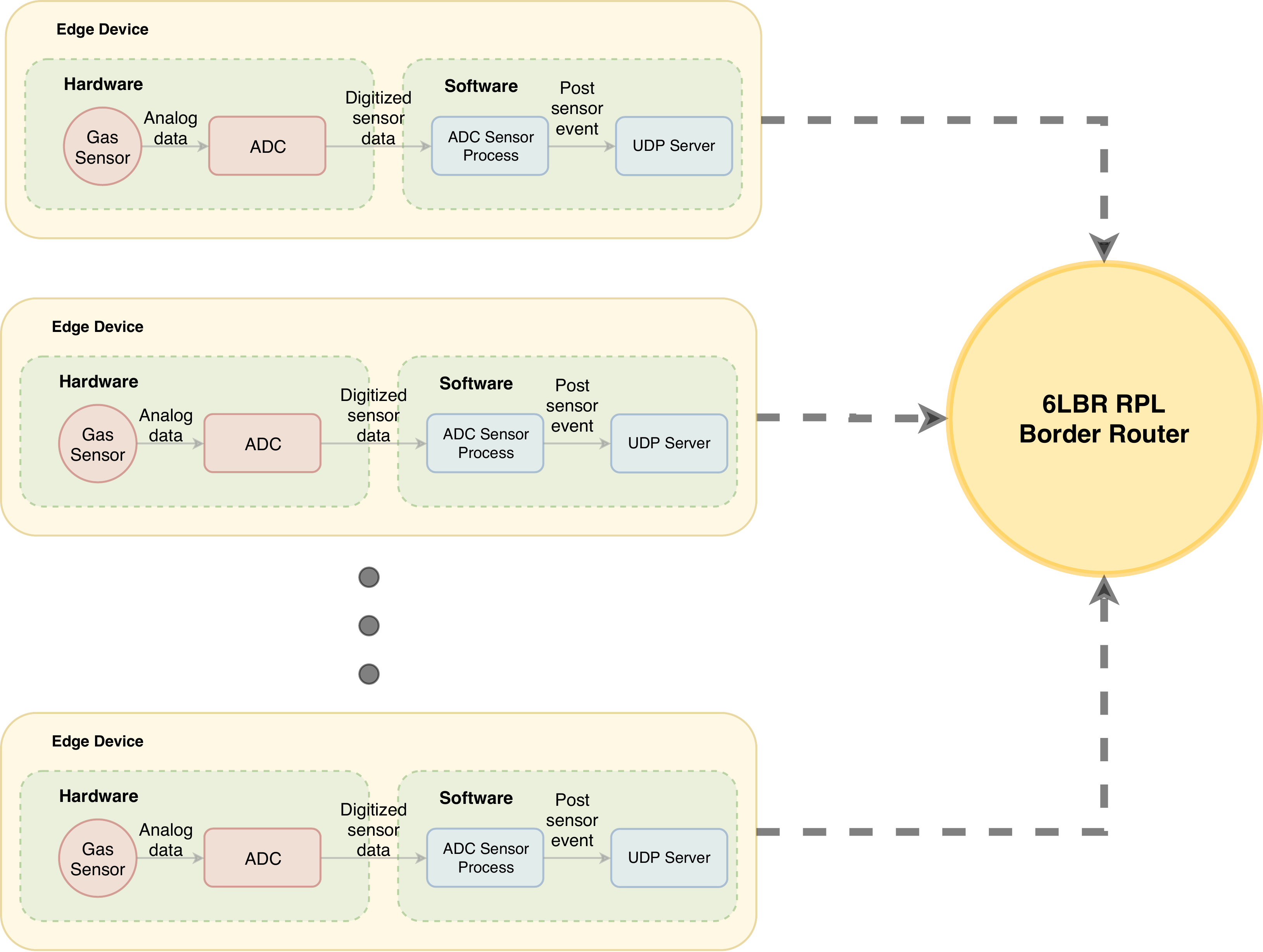}
\vspace{-0.2in}
\caption{6LoWPAN Network of edge devices and border-router}\label{fig:border}
\end{figure}

\textbf{Sensor Data Transmission:} The UDP server process thread collects the data from the process event queue, forms a UDP packet, and sends it to the border router. The UDP servers from different motes send the sensor data over 6LoWPAN to the Border Router (6LBR), as shown in Fig~\ref{fig:border}. 
The 6LBR maintains routing information of all the nodes in the network using RPL, and also forwards data from the 6LoWPAN to a host outside this network, as shown in Fig~\ref{fig:lpwan-tree}. %The two functions of 6LBR are described below

\begin{figure}[t]
\centering
\includegraphics[width=1.0\columnwidth]{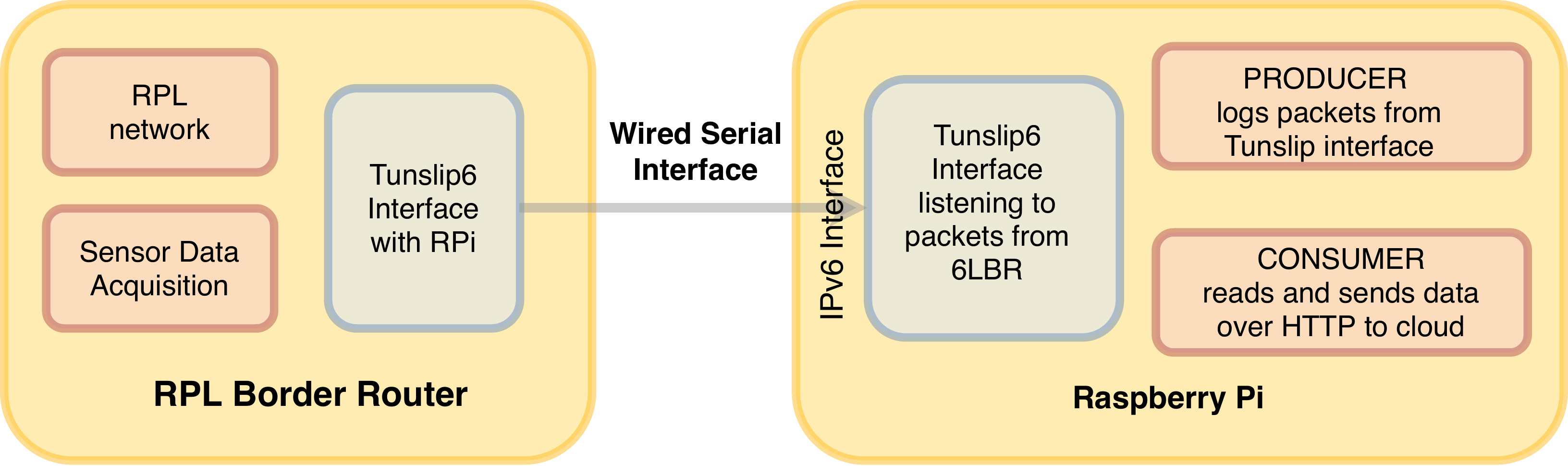}
\vspace{-0.2in}
\caption{Tunslip6 interface between border router and Raspberry Pi}\label{fig:tunslip}
\vspace{-0.1in}
\end{figure}

\textbf{Integration of Raspberry Pi with Border Router:} %RPL implemented on the 6LBR  maintains routing information pertaining to all the nodes in the network. 
One of features of the border router is connecting one network to another. The sensor data from UDP servers are routed to the destination host that is outside the 6LoWPAN, say a Raspberry Pi, through the 6LBR, as follows.
%The data from UDP servers is sent to a host . 

The 6LBR is interfaced with the Pi over a USB serial interface, as shown in Fig~\ref{fig:tunslip}. \emph{Tunslip6} is a tool used to bridge the IP traffic between a host and another network typically a border router, over a serial line. Tunslip6 creates a virtual network interface (tun) on the host and uses the Serial Line Internet Protocol (SLIP)~\cite{rfc1055} to wrap and forward IP packets to and from the other side of the serial line. This interface behaves like any other network interface and supports routing, traffic forwarding etc. We create a tunslip interface at the Pi with an IPv6 address, whose prefix is sent to the 6LBR. The border router in turn broadcasts this prefix to the nodes in its network. All packets addressed to the (Pi) host from any node in the 6LoWPAN will be forwarded from the 6LBR to the host through the tunslip interface.
%e have seen the RPL network so far. The border router is interfaced with Raspberry Pi over USB serial interface. A Tunslip6 interface is created at the Pi with an IPv6 address, whose prefix is sent to the 6LBR.

\begin{figure}[t]
\centering
\includegraphics[width=1.0\columnwidth]{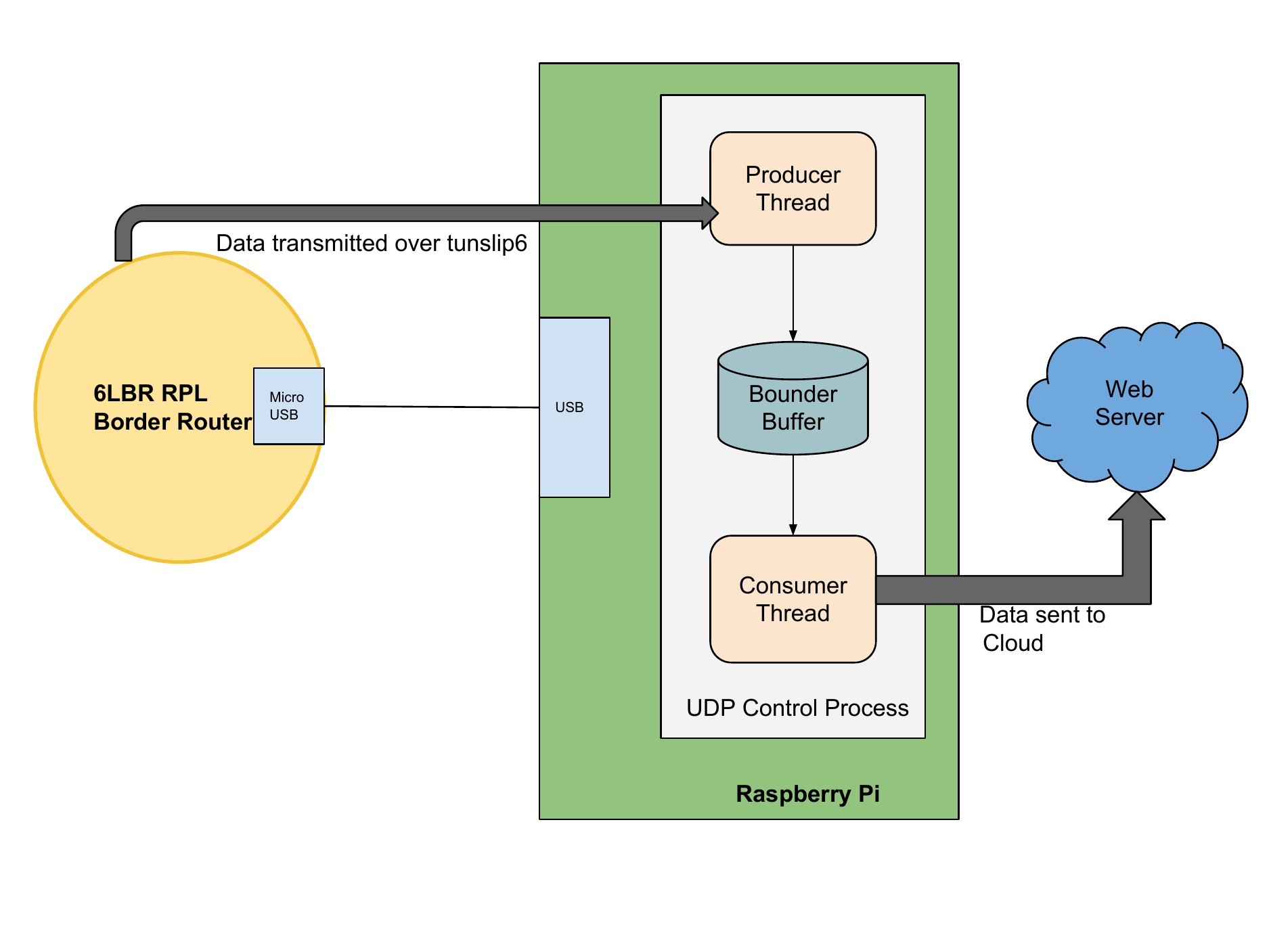}
\vspace{-0.5in}
\caption{Data flow between border router, Raspberry Pi and the cloud}\label{fig:rpi_updated}
\vspace{-0.1in}
\end{figure}

%The border router broadcasts the prefix with which the tunslip6 interface is created to the nodes in its network. All packets addressed to the host will be forwarded from the 6LBR to the host through the tunslip6 interface. 
\textbf{Integration of Raspberry Pi with Cloud:} On the Raspberry Pi side, a \emph{UDP control process} is implemented that listens to the UDP data sent from the UDP process threads on the motes, at port 5000. The host control process on receiving a UDP packet logs the data into a file as well as sends the data to a web server running in the cloud where interesting analytics and visualization is performed on this time series data. This is shown in Fig.~\ref{fig:rpi_updated}. 

The UDP control process uses the \emph{producer-consumer pattern}~\cite{producer-consumer-problem}, with a bounded buffer between the producer and consumer threads whose access is synchronized using a binary semaphore~\cite{binary-semaphore}. The producer thread is continuously listening for data on the tun interface and on receipt of data, it adds it to the buffer. The consumer thread is consuming the data from the buffer (when available) and sends it to the web server running in the cloud. %The schematic diagram showing the interaction of the border router, Raspberry Pi and the cloud is 
The capacity of the buffer is set to withstand 24~hrs of sensor data production without any consumption, in case the network link from the Pi to the cloud or the web-service is down. Also, the data is backed up to the SD card on the Pi, on receipt by the producer so that even if the link is down after 24~hrs, the data is not lost and can be retransmitted from disk once the transient network issue is resolved.

\section{Conclusion}
\label{sec:conclude}

The work presented here is part of a multi-year multi-institution project SATVAM~\footnote{ACKNOWLEDGEMENT: This research was supported under the Research Initiative for Real-time River Water and Air Quality Monitoring program funded by the Department of Science and Technology, Govt. of India (DST) and Intel\textsuperscript{\textregistered}, and administered by the Indo-U.S. Science and Technology Forum (IUSSTF).}. The first stage in this is to build a prototype using commercial off-the-shelf components and evaluate their performance against reference grade equipment. The focus has been on identifying low-powered components which provide flexibility in development and stability in initial field deployments. The next step is to setup a 40 node testbed of the air quality sensors at IIT Kanpur and IIT Bombay. Field calibrations of the sensor data and performance of the monitoring devices (power consumption, reliability)  will be closely monitored. Advanced edge and cloud based analytics will then be developed. The final stage involves a 60 node real-time air quality monitoring network deployed in a city.

% \section*{Acknowledgment}
% This research was supported under the Research Initiative for Real-time River Water and Air Quality Monitoring program funded by the Department of Science and Technology, Govt. of India (DST) and Intel\textsuperscript{\textregistered}, and administered by the Indo-U.S. Science and Technology Forum (IUSSTF).

%\Note{Max 3 pages at this point. 1 more page for references}

\ifCLASSOPTIONcaptionsoff
  \newpage
\fi



\bibliographystyle{IEEEtran}
\bibliography{paper}

\end{document}